\documentclass[twocolumn,aps,prd,preprintnumbers,superscriptaddress,10pt]{revtex4-1}
\usepackage{amsmath,amssymb,amsfonts,mathtools}
\usepackage{epsfig}
\usepackage{graphicx}
\usepackage{slashed}
\usepackage{color}
\usepackage[normalem]{ulem}

\begin{document}

\title{Transverse-Momentum-Dependent Parton Distribution Functions \\ from Large-Momentum Effective Theory}

\author{Xiangdong Ji}
\affiliation{Department of Physics, University of Maryland, College Park, MD 20742, USA}
\affiliation{Tsung-Dao Lee Institute, Shanghai Jiao Tong University, Shanghai 200240, China}

\author{Yizhuang Liu}
\email{yizhuang.liu@sjtu.edu.cn}
\affiliation{Tsung-Dao Lee Institute, Shanghai Jiao Tong University, Shanghai 200240, China}

\author{Yu-Sheng Liu}
\affiliation{Tsung-Dao Lee Institute, Shanghai Jiao Tong University, Shanghai 200240, China}

\date{\today}

\begin{abstract}
We show that transverse-momentum-dependent parton distribution functions (TMDPDFs), important non-perturbative
quantities for describing the properties of hadrons in high-energy scattering processes such as Drell-Yan and semi-inclusive deep-inelastic scattering with observed small transverse momentum, can be obtained from Euclidean QCD calculations in the framework of large-momentum effective theory (LaMET).
We present a LaMET factorization of the Euclidean quasi-TMDPDFs in terms of the physical TMDPDFs and off-light-cone
soft function at leading order in $1/P^z$ expansion, with the perturbative matching coefficient satisfying a renormalization group equation.
We also discuss the implementation in lattice QCD with finite-length gauge links as well as the rapidity-regularization-independent factorization
for Drell-Yan cross section.
\end{abstract}

\maketitle

{\it Introduction.}---High-energy hadron processes involving measuring particles with small transverse momentum ($k_\perp\sim \Lambda_{\rm QCD}$) has been of great interest for particle and nuclear physicists for many years~\cite{Collins:1981uk,Collins:1984kg}.
On the one hand, many productions, such as Drell-Yan process with
transverse momentum $Q_\perp$ measured~\cite{Bacchetta:2019sam,Scimemi:2019cmh}, peak at relatively small $Q_\perp^2\ll Q^2$
where $Q^2$ is the total mass squared of the Drell-Yan pair and cannot be explained entirely with the collinear parton distribution functions (PDFs) and perturbative quantum chromodynamics (QCD).
On the other hand, quarks and gluons confined in a high-energy hadron such as the proton do carry physical transverse momenta which warrant new nonperturbative observables to describe.
The transverse-momentum-dependent parton distribution functions (TMDPDFs) are the simplest extension of the standard textbook PDFs and have inspired a large body of theoretical works~\cite{Collins:1981uk,Collins:1984kg,Collins:1988ig,Ji:2004wu,Ji:2004xq,Collins:2011ca,Collins:2012uy,Becher:2010tm,GarciaEchevarria:2011rb,Echevarria:2012js,Chiu:2012ir}.
However, because TMDPDFs involve the lightcone correlations of quark and gluon fields, there has been little attempt to compute them from lattice QCD (see~\cite{Hagler:2009mb,Musch:2010ka,Musch:2011er,Engelhardt:2015xja,Yoon:2017qzo}
for exploratory studies where the longitudinal momentum $x$-dependence and matching to the continuum definition have not been considered).

The recent development of large-momentum effective theory (LaMET) proposed by one of the authors~\cite{Ji:2013dva,Ji:2014gla} has opened up a possibility of directly calculating TMDPDFs in lattice QCD, a Euclidean formulation of non-perturbative field theory calculations.
The essence of LaMET is simple: while Euclidean lattice does not support modes travelling along the light front (LF) at operators level, it does support on-shell fast moving hadrons which allow extraction of collinear physics
through large-momentum factorization.
However, previous attempts toward a Euclidean formulation of TMDPDFs in the framework of LaMET have not met with complete success because of the presence of the soft contribution~\cite{Ji:2014hxa,Ji:2018hvs,Ebert:2018gzl,Ebert:2019okf}.
The so-called soft functions appearing in various high-energy processes summarize the soft gluon radiation effects of fast moving charged particle.
The soft function in transverse-momentum-dependent processes (TMD soft function) involves two opposite light-like directions and presents a crucial difficulty to implement on Euclidean lattice.
However, the recent progress by the present authors has showed that it can be calculated as the large-velocity-transfer form factor of a fixed-separation color-anticolor source pair and that of a light meson~\cite{Ji:2019sxk}.
Thus the final obstacle to formulate a LaMET calculation for TMDPDFs has been removed.

In this paper, we present essential ingredients for a lattice calculation of TMDPDFs in
a hadron travelling at large momentum.
We start by defining the TMDPDF using the standard LF correlation in dimensional regularization (DR)
and modified minimal-subtraction ($\overline{\rm MS}$) scheme, just as in the collinear
PDF case.
We then introduce the corresponding quasi-TMDPDF using a Euclidean equal-time correlator
at large longitudinal momentum ($P^z$), which captures the same nonperturbative physics
in the $P^z\to \infty$ limit.
We present a LaMET factorization which relates the quasi-TMDPDF
to the physical TMDPDF and off-light-cone soft function in a systematic expansion in $1/P^z$, and derive
a renormalization group equation for the perturbative matching coefficient.
Finally, we discuss subtleties related to the lattice implementation of the quasi-TMDPDF:
finite-length gauge links and self-energy subtraction, as well as UV renormalization.
And we also derive a factorization formula for Drell-Yan cross section directly
in terms of quasi-TMDPDFs which do not involve the light-cone (parton) limit.

\vspace{0.2cm}

{\it Definition of physical TMDPDF.}---Defining a proper TMDPDF that becomes a common standard for both phenomenological data fitting and perturbative QCD calculations turns out be more challenging than that for the collinear PDFs~\cite{Collins:2011zzd}.
The LF divergences from light-like Wilson lines cannot be simply regulated in DR: an extra rapidity
regularization has to be introduced.
The regularization effectively introduces a rapidity cutoff (regulator) for small $k^+$ or large $k^-$ modes
for a hadron travelling in the positive $z$-direction (see definition of $\pm$ component below), and the cutoff dependencies transmute to the rapidity scale dependence $\zeta$.
Many proposals on rapidity regulator have been made in the literature, which result in different ways to isolate the soft physics that does
not cancel for transverse-momentum-dependent processes.
A physics-motivated definition~\cite{Collins:2011ca,Echevarria:2012js,Collins:2012uy} in which
both collinear and (non-cancelling) soft physics are properly combined, called physical TMDPDF, turns out to be
rapidity-regularization-scheme independent (see~\cite{Ebert:2019okf} for a nice summary).

\begin{figure}[htb]
\centering
\includegraphics[width=0.9\columnwidth]{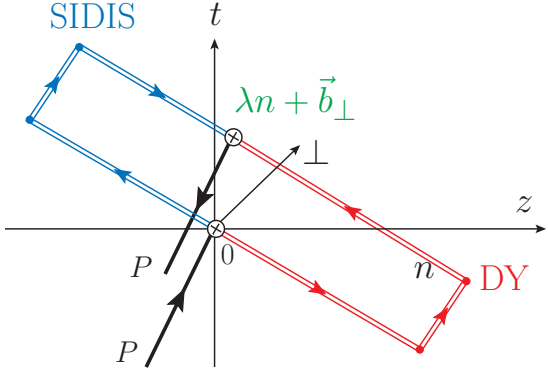}
\caption{The space-time picture of TMDPDF for DY and SIDIS processes. The circled crosses denote the quark-link vertices.}
\label{Fig:tmdpdf}
\end{figure}
To define a physical distribution, we start from the un-subtracted (containing
additional soft modes) distribution function:
\begin{align}
\label{eq:unsubtmd}
& f(x,b_\perp,\mu,\delta^-)= \int \frac{d\lambda}{4\pi}e^{-i\lambda x}\\
&\times\langle P| \bar \psi(\lambda n/2+\vec{b}_\perp)\slashed{n}{\cal W}_{\pm n}
(\lambda n/2+\vec{b}_\perp)|_{\delta^-}\psi(-\lambda n/2) |P\rangle  \,, \nonumber
\end{align}
where $|P\rangle$ is the hadron state labeled by momentum $P=(P^0,0,0,P^z)$, where $P^z$ can be of
any value, in particular, in the rest frame $P^z=0$;
$n^\mu$ is a light-like vector in the minus LF direction $n^\mu=\frac{1}{\sqrt{2}P^+}(1,0,0,-1)$
where the standard LF vector notation $P^\pm = (P^0\pm P^z)/\sqrt{2}$ is used, and $P\cdot n = 1$.
$\lambda$ is the LF coordinate distance whose Fourier-conjugation variable is $x$, the longitudinal momentum fraction of the parton. $b_\perp$ is the transverse separation conjugating to the transverse momentum $k_\perp$. $\mu$ is
an ultra-violate (UV) renormalization scale in $\overline{\rm MS}$ scheme. Finally, ${\cal W}_n(\lambda n+\vec{b}_\perp)$ is the staple-shaped gauge-link of the form
\begin{align}
&{\cal W}_{\pm n}(\xi)|_{\delta}=W^{\dagger}_{\pm n}(\xi)W_{\perp}W_{\pm n}(-\xi \cdot p\,n)\big|_{\delta} \label{eq:staplen} \ , \\
&W_{\pm n}(\xi)= {\cal P}\exp\left[-ig\int_{0}^{\pm\infty} ds\,n\cdot A(\xi+n s)\right] \ ,
\end{align}
where $A^\mu$ is the gauge potential in QCD, $W_\perp$ is the gauge link in the transverse direction connecting the
end-points at the infinity of $W_{\pm n}$~\cite{Belitsky:2002sm}, and ${\cal P}$ stands for path ordering.
The subscript $\delta$ in $|_\delta$ indicates that the gauge links along the LF direction have LF divergences
which cannot be regulated by the usual DR, and a new LF regulator is needed (we generically call it $\delta$ although
there are two general types: off-light-cone and on-light-cone)~\cite{Collins:2012uy,Becher:2010tm,GarciaEchevarria:2011rb,Echevarria:2012js,Chiu:2012ir}. Because
of this new regulator, the quark bilinear operator is not boost-invariant and the
resulting distribution is no longer independent of $P^z$.
The $\pm$ sign choice in $\cal W$ is process dependent: For the Drell-Yan process one should choose negative sign while for the semi-inclusive deep inelastic scattering (SIDIS) one should choose positive sign~\cite{Collins:2004nx,Collins:2011zzd}.
For unpolarized TMDPDF interested in here, two choices are equivalent. It is straightforward to generalize the discussions to spin dependence as well as gluons. We have shown
the path of the gauge link in Fig.~\ref{Fig:tmdpdf}.

The physical TMDPDF involves a combination of the above distribution with the square root of the soft contribution $S(b_\perp,\mu,\delta e^{2y_n},\delta)$ divided~\cite{Collins:2012uy},
\begin{align}\label{eq:physical_TMD}
f^{\rm TMD}(x,b_\perp,\mu,\zeta)=\lim_{\delta\rightarrow 0}\frac{f(x,b_\perp,\mu,\delta)}{\sqrt{S(b_\perp,\mu,\delta e^{2y_n},\delta)}}
\end{align}
where the LF soft function is defined by two Wilson lines in conjugate LF directions~\cite{Collins:2011zzd},
\begin{align}\label{eq:soft}
& S(b_\perp,\mu,\delta^+,\delta^-)=\frac{1}{N_c} \\
&\times{\rm Tr}\langle 0|{\cal \bar T}[W_p(\vec{b}_\perp)|_{\delta^+}W_n^{\dagger}(\vec{b}_\perp)|_{\delta^-}]{\cal T} [W_n(0)|_{\delta^-}W^{\dagger}_p(0)|_{\delta^+}]|0\rangle\nonumber \,,
\end{align}
with $N_c=3$ is the number of color and the trace is over the color space, $p^\mu=\frac{P^+}{\sqrt{2}}(1,0,0,1)$ is in the plus LF direction (although $p$ and $n$ carry mass dimensions, the Wilson line depends only on the direction of $p$ or $n$, not their dimensions as can be shown by a recalling of the path parameter $s$). ${\cal T}$ and $\bar{\cal T}$ stands for time and anti-time orderings, respectively. $\delta^\pm$ is LF regulators in the LF $\pm$ directions. In Eq. (\ref{eq:physical_TMD}), $\delta \equiv \delta^-$, and
$\delta^+ \equiv \delta e^{2y_n}$. In the physical distribution, only combination called Collins-Soper scale
$\zeta=2(xP^+)^2e^{2y_n}$ appears.

\begin{figure}[htb]
\centering
\includegraphics[width=0.9\columnwidth]{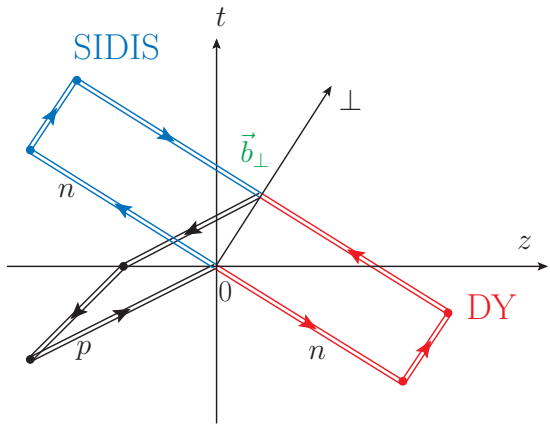}
\caption{The soft function $S(b_\perp,\mu,\delta^+,\delta^-)$ as space-time Wilson-loop arising in the
factorization of DY and SIDIS processes.}
\label{Fig:soft1}
\end{figure}

Finally, the renormalization group equation for the TMDPDF is~\cite{Becher:2010tm,Chiu:2012ir},
\begin{align}
\mu\frac{d}{d\mu} \ln {f^{\rm TMD}}(x,b_\perp,\mu,\zeta)=\Gamma_{\rm cusp}(\alpha_s)\ln\frac{\mu^2}{\zeta} -2\gamma_H(\alpha_s) \ ,
\end{align}
where $\gamma_{H} $ is the hard anomalous dimension, and $\Gamma_{\rm cusp}$ is the universal light-like cusp anomalous dimension~\cite{Korchemsky:1987wg,Grozin:2015kna}. The rapidity dependence obeys Collins-Soper
equation~\cite{Collins:2011zzd},
\begin{align}
2\zeta \frac{d}{d\zeta} \ln f^{\rm TMD}(x,b_\perp,\mu,\zeta)= K(b_\perp,\mu)
\end{align}
where $K$ is the Collins-Soper evolution kernel which is non-perturbative
for large $b_\perp$. The renormalization group equation for $K$ is also controlled
by the same cusp anomalous dimension $\Gamma_{\rm cusp}$, $\frac{d\ln K}{d\ln \mu} = -2\Gamma_{\rm cusp}$.

\vspace{0.2cm}

{\it Quasi-TMDPDF and off-light-cone soft function.}---
To calculate TMDPDFs in Euclidean space, we need to have a new representation
of partons in which the hadron state $P^\mu=(P^0,0,0,P^z)$ must be in the infinite-momentum frame (IMF), $P^z\to\infty$~\cite{Ji:2014gla,Ji:2020byp}. This is
in fact the old-fashioned, Feynman's IMF approach to partons~\cite{Feynman:1973xc}.
In the large momentum limit, different operators (for example, $\bar\psi\gamma^0\psi$ and $\bar\psi\gamma^z\psi$) between infinite-momentum states can produce
the same parton physics, a phenomena called universality~\cite{Ji:2014gla,Hatta:2013gta}.
A LaMET theory for partons is based on the above Feynman's picture. It
consists of approximating the matrix elements at the infinite $P^z$ by a finite but large $P^z\gg M$,
where $M$ is the hadron mass, and expanding the result around $P^z=\infty$. [Note that the matrix elements
are calculated with the UV cut-off $\Lambda_{\rm UV}$ such as the largest momentum supported by
a lattice must be much larger than $P^z$.] They are finally matched
to the standard partonic observables defined using LF correlators, such as Eq.~(\ref{eq:unsubtmd}) and Eq.~(\ref{eq:soft}). This last step is very similar to the heavy-quark effective theory~\cite{Manohar:2000dt}, where the finite
mass of the quark $m_Q$ is replaced by an infinite heavy quark $m_Q=\infty$.

\begin{figure}[htb]
\centering
\includegraphics[width=0.8\columnwidth]{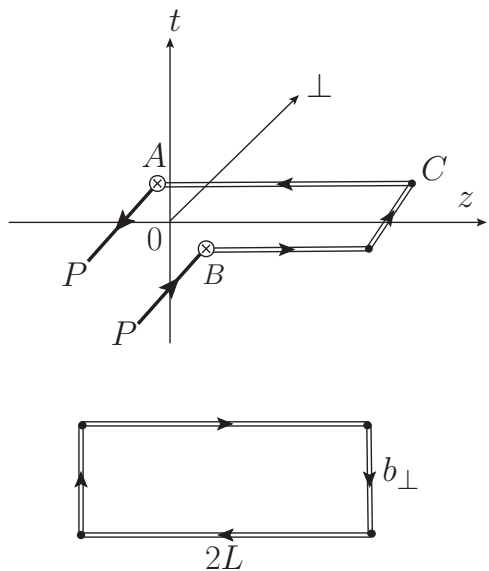}.
\caption{The space-time pictures for quasi-TMDPDF (upper) and the Wilson-loop $Z_E(2L,b_\perp,\mu,0)$ (lower). The
coordinates of the points are $A=\lambda n_z/2+\vec{b}_\perp/2$, $B=-\lambda n_z/2-\vec{b}_\perp/2$ and $C=Ln_z+{\vec b}_\perp$. The crosses denote the quark-link vertices.}
\label{fig:quasiTMD}
\end{figure}

According to LaMET, calculation of TMDPDF starts with the physical
momentum distributions, or the Fourier space correlations of quarks (and gluons) at
a hadron with momentum $P^z$ ~\cite{Ji:2013dva,Ji:2014gla,Ji:2014hxa,Ji:2018hvs,Ebert:2018gzl,Ebert:2019okf},
\begin{align}\label{eq:quasi_tmd}
& \tilde f(\lambda ,b_\perp,\mu,P^z) \\
&=\! \lim_{L \rightarrow \infty}  \frac{\langle P| \bar \psi\big(\frac{\lambda n_z }{2}\!+\!\vec{b}_\perp\big)\gamma^z{\cal W}_{z}(\frac{\lambda n_z}{2}\!+\!\vec{b}_\perp;L)\psi\big(\!-\!\frac{\lambda n_z}{2}\big) |P\rangle}{\sqrt{Z_E(2L,b_\perp,\mu)}} \ , \nonumber
\end{align}
where the $\overline{\rm MS}$ renormalization is implied, and
\begin{align}\label{eq:staplez}
&{\cal W}_z(\xi;L)=W^{\dagger}_{z}(\xi; L)W_{\perp}W_{z}(-\xi^zn_z;L)  \ ,\\
&W_{z}(\xi;L)= {\cal P}{\rm exp}\Big[-ig\int_{\xi^z}^{L} ds\, n_z\cdot A(\vec{\xi}_\perp\!+\!n_z s)\Big] \ ,
 \end{align}
Here $\xi^z=-\xi\cdot n_z$, $n_z^\mu=(0,0,0,1)$. $W_{\perp}$ is inserted at $z=L$ to maintain gauge invariance.
Because of universality, $\gamma^z$ can be replaced by $\gamma^0$ in Eq.~(\ref{eq:quasi_tmd}), or any linear
combination of them except $\gamma^0-\gamma^z$.
The $P^z$ dependence of the quasi-TMDPDF originates from non-invariance of the Euclidean bilocal operator under Lorentz boost.
The detailed depiction of the operator with Wilson lines is shown in the top panel in Fig. \ref{fig:quasiTMD}, which
is time-independent. $\sqrt{Z_E(2L,b_\perp,\mu,0)}$ is the square root of the vacuum expectation value of a flat rectangular Euclidean Wilson-loop (shown in the bottom panel of Fig. \ref{fig:quasiTMD}) along the $n_z$ direction with length $2L$ and width $b_\perp$:
\begin{align}\label{eq:Z_E}
Z_E(2L,b_\perp,\mu)=\frac{1}{N_c}{\rm Tr}\langle 0|W_{\perp}{\cal W}_z(\vec{b}_\perp;2L)|0\rangle \ .
\end{align}
The self-interactions of gauge links are subtracted using $\sqrt{Z_{E}}$ in order to remove the pinch-pole singularity~\cite{Collins:2008ht,Ji:2019sxk}, which will be discussed in more detail later.
The Fourier transformation of the correlator $\tilde f(\lambda ,b_\perp,\mu,P^z)$ in Eq. (\ref{eq:quasi_tmd})
with respect to $\lambda$ defines the momentum distribution $\tilde f(x,b_\perp,\mu,\zeta_z)$, where $\zeta_z=(2xP^z)^2$ is the Collins-Soper scale,
which we call a quasi-TMDPDF in the large momentum limit because it now can be matched to parton physics.

The quasi-TMDPDF does not have any rapidity divergences because it does not
involve any light-like separations in quantum fields. For a physical process, there is no rapidity divergence,
and the large $k^\pm$ modes are naturally bounded by the physical hadron momentum, which
naturally plays the role of a rapidity regulator. Therefore,
the quasi-TMDPDF can be regarded as the definition of TMDPDF with $P^z$ as the rapidity regulator.
In fact, if one un-boosts the hadron with the large momentum $P^z$ and applies
the boost operator to the equal-time correlator in Eq.~(\ref{eq:quasi_tmd}) instead,
one finds a quasi-LF correlator which becomes exactly the LF correlator in the $P^z\to \infty$ limit~\cite{Ji:2014gla}.
The size of $P^z$ determines the angle between the boosted Wilson line and the light
front, which justifies $P^z$ as an off-light-cone regulator~\cite{Collins:1981uk,Ji:2004wu}.

According to the argument above, one would expect that the change of the quasi-TMDPDF with respect to
the hadron momentum is also equivalent to the rapidity evolution extracted from the cutoff dependence.
Recall that the momentum evolution for the quasi-PDFs in the collinear factorization reproduces
the well-known DGLAP evolutions~\cite{Ji:2014gla,Ji:2020ect}.
Indeed, the $\zeta_z$ evolution of $\tilde f(x,\zeta_z,b_\perp,\mu)$ is
exactly the Collins-Soper evolution~\cite{Collins:1981uk,Ji:2014hxa}
\begin{align}\label{eq:quasievol}
2\zeta_z \frac{d}{d\zeta_z} \ln \tilde f(x,b_\perp,\mu,\zeta_z)= K(b_\perp,\mu)+G(\zeta_z,\mu)
\end{align}
where $K$ is independent of rapidity regularization scheme. Therefore,
$K$ can be calculated non-perturbatively from $P^z$-dependence of quasi TMDPDF~\cite{Ebert:2018gzl,Shanahan:2019zcq,Zhang:2020dbb}.
Compared with Eq. (7), the extra term $G(\zeta_z,\mu)$ exists only in the off-light-cone scheme. It is due to the presence of the hard scale $\zeta_z=\frac{(2xP\cdot n_z)^2}{|n_z^2|}$, a Lorentz invariant combination of the parton momentum $xP$ and the off-light-cone direction vector $n_z$. Therefore, it is perturbative in nature.
At one-loop level, one has~\cite{Collins:1981uk,Ji:2014hxa},
\begin{align}
K^{(1)}(b_\perp,\mu)&=-\frac{\alpha_sC_F}{\pi}\ln \frac{\mu^2 b_\perp^2}{4e^{-2\gamma_E}}\, ,\\
G^{(1)}(\zeta_z,\mu)&=\frac{\alpha_sC_F}{\pi}\left(1-\ln\frac{\zeta_z}{\mu^2}\right) \, .
\end{align}
Because $\tilde f$ has a simple UV renormalization, that the combination $K+G$ is renormalization group invariant~\cite{Collins:1981uk}, and thus $\frac{d\ln G}{d\ln \mu} = 2\Gamma_{\rm cusp}$.

The quasi-TMDPDF satisfies a renormalization group equation in DR ~\cite{Collins:1981uk,Ji:2004wu},
\begin{align}\label{eq:quasiRGE}
\mu\frac{d}{d\mu}\ln {\tilde f}(x,b_\perp,\mu,\zeta_z) = 2\gamma_F(\alpha_s)
\end{align}
where $\gamma_F$ is the anomalous dimension of the quark field in $A^z=0$ gauge, or the anomalous dimension for the heavy-light quark current~\cite{Falk:1990yz,Ji:1991pr,Braun:2020ymy}. The anomalous dimension of
the Wilson-line cusps in the numerator of Eq.~(\ref{eq:quasi_tmd}) has been subtracted by $\sqrt{Z_E}$,
and hence does not appear.


Since the physical combination of un-subtracted TMDPDF and the soft function defined in the same rapidity
regularization scheme is scheme-independent, a calculation of physical
TMDPDF in LaMET requires not only the un-subtracted
$\tilde f(x,b_\perp,\mu,\zeta_z)$, but also the soft function $S$ in the same off-light-cone scheme.
Finding a Euclidean formulation of this soft function is critical
for the success of the quasi TMDPDF approach, although in a small number of
applications, this soft function can eliminated through ratios.

According to Ref. \cite{Ji:2019sxk}, the required soft function can be obtained as the form factor of a heavy quark-antiquark system:
\begin{align} \label{eq:soft_HQET}
S(b_\perp,\mu,Y,Y') = {}_{v'}\langle Q\bar Q|J(v,v',\vec b_\perp)|Q\bar Q\rangle_v
\end{align}
where $v$ and $v'$ are four-velocities slightly off lightcone, and $Y$ is the rapidity corresponding to $v$.
The state $|Q\bar Q\rangle_v$ is a pair of color-anticolor source separated by distance $b_\perp$
and travelling with velocity $v$.
The ``transition current'' $J$ is a product of two equal-time local operators,
\begin{align}
J(v,v',\vec{b}_\perp)= \bar Q^\dagger_{v'}(\vec{b}_\perp)\bar Q_v(\vec{b}_\perp)Q_{v'}^\dagger(0)Q_v(0) \ .
\end{align}
where $Q_v$ and $\bar Q_v$ are quark and anti-quark fields in the fundamental and anti-fundamental representations, respectively. The same soft factor can also be extracted from a form factor and quasi-TMD wave function
of a light meson~\cite{Ji:2019sxk}, and a first calculation on lattice has appeared recently~\cite{Zhang:2020dbb}.

At large rapidities, one can write $S(b_\perp,\mu,Y,Y')$ as~\cite{Collins:2011zzd}:
\begin{align}
S(b_\perp,\mu,Y,Y')=e^{(Y+Y')K(b_\perp,\mu)}S_r^{-1}(b_\perp,\mu) + ...
\end{align}
where $K(b_\perp,\mu)$ is the same Colins-Soper evolution kernel discussed above.
The reduced soft function $S_r(b_\perp,\mu)$ is rapidity-independent, but
rapidity-regulator dependent. Its renormalization scale dependence is~\cite{Ji:2019sxk}:
\begin{align}
&\mu^2 \frac{d}{d\mu^2} \ln S(b_\perp,\mu,Y,Y')\nonumber \\
&\qquad\qquad=-(Y+Y')\Gamma_{\rm cusp}(\alpha_s)-\Gamma_S(\alpha_s)\,,
\end{align}
where $\Gamma_S$ is the constant part of the cusp-anomalous dimension at large hyperbolic cusp angle $Y+Y'$ for the off-light-cone soft function~\cite{Collins:2011zzd}.
Thus we find,
\begin{align}
\mu^2 \frac{d}{d\mu^2} \ln S_r(b_\perp,\mu)=\Gamma_S(\alpha_s) \,,
\end{align}
Note that the scale evolution of the TMD soft function is controlled by cusp anomalous dimension at large hyperbolic angle rather than a circular angle.

At one-loop for small $b_\perp$, the soft function is~\cite{Ji:2004wu,Ebert:2019okf}
\begin{align}
S^{(1)}(b_\perp,\mu,Y,Y')=\frac{\alpha_sC_F}{2\pi}\big[2-2(Y+Y')\big]\ln\frac{\mu^2b_\perp^2}{4e^{-2\gamma_E}} \,.
\end{align}
Perturbative result beyond one loop is unknown so far.

\vspace{0.2cm}

{\it Factorization of quasi-TMDPDF and evolution of matching kernel}---
According to the above discussion, the difference between quasi-TMDPDF $\tilde f$
and the un-subtracted TMDPDF $f$ is mainly caused by the difference in
rapidity regularization and the ordering between UV regularization and
large-momentum limit. This latter point needs some clarification:
While the quasi-TMDPDF is obtained at $\Lambda_{\rm UV} \gg P^z$,
the LF definition of TMDPDF uses the limit $P^z\to \infty$ before $\Lambda_{\rm UV}$ is applied.
Therefore, the infrared physics in both $\tilde f$ and $f$ must be the same up to soft physics defined
by the rapidity regulator. It is then natural that the physical combinations
of the un-subtracted TMDPDFs and soft functions in different rapidity
regularization schemes differ only by
a perturbative factor. Hence one can write down
the following factorization or matching formula,
\begin{align}\label{eq:factorization}
&\tilde f(x,b_\perp,\mu,\zeta_z)\sqrt{S_r(b_\perp,\mu)}\nonumber \\
&=H\left(\frac{\zeta_z}{\mu^2}\right) e^{K(b_\perp,\mu)\ln (\frac{\zeta_z}{\zeta})}f^{\rm TMD}(x,b_\perp,\mu,\zeta)+...\ ,
\end{align}
where the power-corrections denoted by the $...$ term are of order ${\cal O}\left(\Lambda_{\rm QCD}^2/\zeta_z,M^2/\zeta_z,1/(b^2_\perp\zeta_z) \right)$. The factor $H$ is the perturbative matching kernel, which is a function of $\zeta_z/\mu^2=(2xP^z)^2/\mu^2$.

\begin{figure}[t]
\includegraphics[width=0.9\columnwidth]{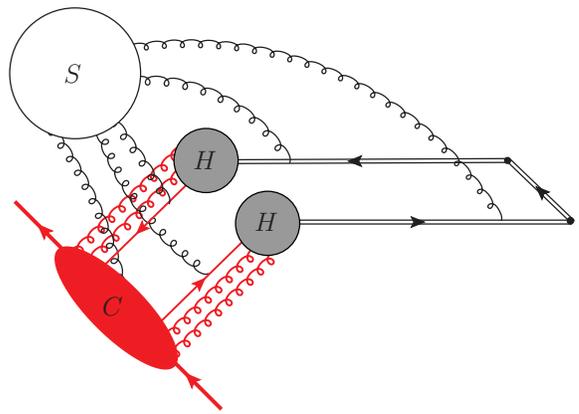}
\caption{The leading regions of the quasi-TMDPDF where $C$ is the collinear sub-diagram, $S$ is the soft sub-diagram and $H$'s are hard sub-diagrams. The two hard cores are not connected with each other, and as a result, the momentum fraction of the quasi-TMDPDF receives contribution only from collinear modes and there is no convolution in the matching formula. }
\label{fig:quasireduce}
\end{figure}

Here we provide a sketch of mathematical proof for Eq.~(\ref{eq:factorization}), leaving details to a future publication~\cite{future}. Following the standard leading-region analysis and power-counting argument~\cite{Collins:2011zzd}, with minor modifications to include the staple-shaped gauge-links, one can show that the leading regions of the quasi-TMDPDF have the form depicted in Fig.~\ref{fig:quasireduce}. The collinear, soft and hard sub-diagrams are responsible for collinear, soft and ultraviolet (or hard) contributions, respectively.
In particular, the collinear contributions are exactly the same as those in the physical TMDPDF
defined by LF correlators, obtained by taking $P^z\to\infty$. The off-light-cone soft function naturally
appear to capture the effect of soft radiations between the fast-moving color-charges and the
staple-shaped gauge-links. The natural
hard scale of the hard sub-diagram is provided by $\zeta_z$, which arises from the Lorentz-invariant
combination of the momentum of collinear modes and the direction $n_z$ in the static operator.
At large $b_\perp \gg \frac{1}{xP^z}$ or small $k_\perp \ll xP^z$, hard contributions are confined within vicinities of the quark-link vertices around $0$ and $b_\perp$, and any hard momenta flowing between $0$ and $b_\perp$ will cause additional power-suppressions in $\frac{1}{P^z}$. As a result,  the momentum fraction in the quasi-TMDPDF
receives contribution only from collinear modes and there is no convolution in the matching formula Eq.~(\ref{eq:factorization}). This is in sharp contrast to the $k_\perp$ integrated quasi-PDFs, in which hard momenta are allowed to flow between the vertices when $k_\perp$ is comparable to $P^z$. This is also reflected by
that the factorization formula for Drell-Yan process at small $Q_\perp$ involves no convolution, while the inclusive Drell-Yan process does.

Given the leading-region analysis, applications of the standard Ward-identity argument~\cite{Collins:2011zzd} will leads to the factorization formula similar to that in Eq.~(\ref{eq:factorization}). The soft function
required for defining $f^{\rm TMD}$ differs from the off-light-cone soft function in $\tilde f$, although
their rapidity-dependent part is the same. The rapidity-independent parts differ exactly by
the reduced soft function $S_r$, which appears in the final matching formula Eq.~(\ref{eq:factorization})~\cite{Ji:2014gla}.
The exponential involving the Collins-Soper kernel can be obtained by combining all the rapidity dependent factors and can be justified in the following way. From the momentum evolution equation in Eq.~(\ref{eq:quasievol}), there are logarithms of the form $K(b_\perp,\mu)\ln \frac{\zeta_z}{\mu^2} $. To match to the TMDPDF at arbitrary $\zeta$, a factor $\exp\big[\ln (\frac{\zeta_z}{\zeta})K(b_\perp,\mu)\big]$ is required to compensate the difference. Finally, the hard contributions are captured by the hard kernel $H$, which depends on the natural hard scale $\zeta_z$ and the renormalization scale $\mu$. They are closely related to the perturbative $G$ term in Eq.~(\ref{eq:quasievol}). By comparing Eqs.~(\ref{eq:factorization}) and~(\ref{eq:quasievol}), we have the constraint, $2\frac{d\ln H}{d\ln \zeta_z}=G$.

Exploring a first-principle calculation of the Drell-Yan cross section at large $b_\perp$
or small $q_\perp$ with proper UV and IR matching has started
from Ref. \cite{Ji:2014hxa}. The main issue has been about defining
an appropriate soft-function subtraction in Euclidean space. In \cite{Ji:2014hxa}, a bent
Euclidean Wilson loop was used, and the rapidity equation for the unsubtracted
quasi-TMDPDF and non-perturbative calculation of the evolution kernel
through $P^z$ dependence were suggested. Matching was made directly to the experimental
cross section at the scale $\zeta_z$.
In Ref. \cite{Ji:2018hvs}, a flat Wilson loop was proposed as the soft function for subtraction,
and a perturbative matching between the subtracted quasi-TMDPDF and $f^{\rm TMD}$ at the same
rapidity ($\zeta_z=\zeta$) was proposed. In Refs.~\cite{Ebert:2018gzl,Ebert:2019okf}, an un-subtracted
factorization formula similar to Eq.~(\ref{eq:factorization}) was first proposed, which contains both an exponential of
non-perturbative Collins-Soper kernel due to the rapidity difference between $\zeta$ and $\zeta_z$,
and an explicit soft factor $g_S(b_\perp)$. At one-loop level, $g_S$ was identified
as the bent Wilson-loop in \cite{Ji:2014hxa}. However, all of these have not achieved the goal
of obtaining the physical TMDPDF from the correct lattice-calculable non-perturbative quantities
augmented with perturbative corrections. Although, the bent soft functions in Ref.~\cite{Ji:2014hxa,Ebert:2019okf} give the correct cusp anomalous dimension at one-loop level, in general it differs from $S_r$ beyond one-loop correction, which be can be shown by comparing the difference of the corresponding anomalous dimensions ~\cite{Korchemsky:1987wg}.
With the Euclidean soft function in Ref. \cite{Ji:2019sxk}, the above factorization formula
allows this goal to be achieved.

From the equations above, particularly, from the properties of the specific soft function
in the off-light-cone scheme,
the matching kernel $H(\zeta_z/\mu^2)$ satisfies the following renormalization group equation:
\begin{align}
\mu\frac{d}{d\mu} \ln H\left(\frac{\zeta_z}{\mu^2}\right)=\Gamma_{\rm cusp} \ln \frac{\zeta_z}{\mu^2} + \gamma_C
\end{align}
where $\gamma_C=2\gamma_F + \Gamma_S +2\gamma_H $.
The general solution to the renormalization group equation reads
\begin{align}\label{eq:C_renormalization}
&H\left(\alpha_s(\mu),\frac{\zeta_z}{\mu^2}\right)=H\left(\alpha_s(\sqrt{\zeta_z}), 1\right)\\
&\times\exp\left\{\int_{\sqrt{\zeta_z}}^{\mu}\frac{d\mu'}{\mu'}\left[\Gamma_{\rm cusp}(\alpha_s(\mu')) \ln \frac{\zeta_z}{\mu^{'^2}}+\gamma_C\big(\alpha_s(\mu')\big)\right]\right\}\nonumber\,.
\end{align}
This equation allows the determination of the logarithmic structure for $H$ to all orders in perturbation theory, up to unknown constants related to the initial condition $H(\alpha_s,1)$.
The $\Gamma_{\rm cusp},\gamma_F,\gamma_H$ are all rapidity-regulator independent
and have been calculated to three-loops~\cite{Chetyrkin:2003vi,Luo:2019hmp,Grozin:2015kna}, while $\Gamma_S$
is defined in the off-light-cone scheme and has been worked out to two-loop level~\cite{Grozin:2015kna}.

At one-loop level, the quasi-TMDPDF and un-subtracted TMDPDF at small $b_\perp$ can be found in Refs.~\cite{Ji:2018hvs,Ebert:2019okf}.
Thus the one-loop matching coefficient reads
\begin{align}
\ln H^{(1)}\left(\alpha_s,\frac{\zeta_z}{\mu^2}\right) =\alpha_s\left[ c_1+\frac{C_F}{2\pi}\left(\ln\frac{\zeta_z}{\mu^2}-\frac{1}{2}\ln^2\frac{\zeta_z}{\mu^2}\right)\right]
\end{align}
where $c_1=\frac{C_F}{2\pi}\left(-2+\frac{\pi^2}{12}\right)$ is determined by perturbation theory at one-loop level~\cite{Ji:2018hvs,Ebert:2019okf}.
We also anticipate that the two-loop matching kernel is of the form
\begin{align}
&\ln H^{(2)}\left(\alpha_s,\frac{\zeta_z}{\mu^2}\right)=\alpha_s^2\Bigg\{c_2-\frac{1}{2}\left(\gamma^{(2)}_C-\beta_0 c_1\right)\ln\frac{\zeta_z}{\mu^2}\nonumber\\
&-\frac{1}{4}\left(\Gamma^{(2)}_{\rm cusp}-\frac{\beta_0 C_F}{2\pi}\right)\ln^2\frac{\zeta_z}{\mu^2}-\frac{\beta_0C_F}{24\pi}\ln^3\frac{\zeta_z}{\mu^2} \Bigg\}
\end{align}
where $\beta_0=-\frac{1}{2\pi}\left(\frac{11}{3}C_A-\frac{4}{3}N_f T_F\right)$ is the coefficient of one-loop $\beta$-function, $\Gamma^{(2)}_{\rm cusp}=C_FC_A(-\frac{1}{12}+\frac{67}{36\pi^2})-\frac{5C_FN_f}{18\pi^2}$ is the two-loop light-like cusp anomalous dimension and $\gamma_C^{(2)}=\frac{1}{16\pi^2}\left(a_1C_FC_A+a_2C_F^2+a_3C_FN_f\right)$ with $a_1=60 \zeta_3-\frac{19\pi ^2}{3}+\frac{68}{27}$, $a_2=-48 \zeta_3+\frac{28 \pi ^2}{3}-8$ and $a_3=\frac{2\pi ^2}{3}-\frac{62}{27}$. $c_2$ is a constant to be determined in perturbation theory at two-loop level.

\vspace{0.2cm}

{\it Discussions and Conclusion.}---To implement a calculation of the quasi distribution on lattice, a few comments are
in order. First, the Euclidean quasi-TMDPDF $\tilde f(x,b_\perp,\mu,\zeta_z)$ shall now be calculated with a finite length $L$ of the staple shaped gauge link and closed with a transverse gauge link with width $b_\perp$~\cite{Musch:2010ka,Ji:2018hvs}.
In large $L$ limit, there are pinch-pole singularities which is responsible for the heavy-quark potential term $e^{-L V(b_\perp)}$.
The gauge link self-energy contains linear divergence~\cite{Dotsenko:1979wb,Brandt:1981kf}. The quark bilinears with the gauge links
can be renormalized multiplicatively as shown in Ref.~\cite{Ishikawa:2017faj,Ji:2017oey}.
Furthermore, there are additional cusp divergences at junctions of the $z$ direction and transverse gauge links.
These extra complications can be eliminated by performing a subtraction using $\sqrt{Z_E(2L,b_\perp)}$ defined by a square root of the vacuum expectation of a spacial rectangular Wilson loop with length $2L$ and width $b_\perp$.
The same type of subtractions applies to the soft function as well.
The na\"{i}ve soft function in the off-light-cone scheme, defined similar to Eq.~(\ref{eq:soft}), also contains two staples along $v$ and $v'$ directions.
Each of the staple also contributes to additional cusp divergences and a divergent time evolution factor associated to the color-anticolor pair.
As a result, two rectangular Wilson loops along these directions are required to remove those singularities.
After the subtraction, the soft function is equal to the form factor in Eq.~(\ref{eq:soft_HQET}).
Second, the correlation function $\tilde f(z,b_\perp, a,P^z)$ on lattice contains no uncancelled linear divergence in the lattice spacing $a$. The logarithmic divergences in $a$ only appear as a $z$-independent overall constant. Therefore,
the matching to $\tilde f(z,b_\perp, \mu,P^z)$ in $\overline{\rm MS}$ scheme is straightforward.
The operator mixing on the finite-spacing lattice has first been explored in Ref. \cite{Shanahan:2019zcq}.

Finally, our formalism allows the factorization of the DY cross section in terms of the quasi-TMDPDF and soft function. Starting from the factorization formula using standard TMDPDFs in the form $\sigma f^{\rm TMD}f^{\rm TMD}$ (Eq. (1) in Ref.~\cite{Collins:2012uy}). One then rewrite the expression in terms of the quasi-TMDPDFs and
reduced soft function with the help of the matching formula Eq.~(\ref{eq:factorization}).
One then arrives at \cite{Ji:2019sxk}
\begin{align}\label{eq:sigma_DY}
\frac{d\sigma_{\rm DY}}{d^2\vec b_\perp\, dx dx'}&=\hat\sigma(x,x',\mu,\zeta_z,\zeta_z')S_r(b_\perp,\mu)\\
&\times\tilde f(x,b_\perp,\mu,\zeta_z)\tilde f(x',b_\perp,\mu,\zeta_z')+ ...\nonumber
\end{align}
where $\hat\sigma$ is the hard kernel, which is related to the known hard kernel $\sigma(Q^2/\mu^2)$~\cite{Moch:2005tm,Baikov:2009bg} for standard TMD-factorization of DY-process through the relation $\sigma(Q^2/\mu^2)= H(\zeta_z/\mu^2)H(\zeta_z'/\mu^2)\hat\sigma(x,x',\mu,\zeta_z,\zeta_z')$. The omitted term $...$ denotes power-corrections that are responsible for the breakdown of TMD-factorization at small $b_\perp$ or large $Q_\perp$. The Collins-Soper kernel contributions in the matching formula cancel out at $\zeta_z\zeta_z'=\zeta^2=Q^4$. For practical predication of DY cross section from current lattice simulations, it is more feasible to calculate the quasi-TMDPDFs at smaller $\zeta_z$ and evolve them to the experiment scale using the Collins-Soper kernel extracted from ratios of quasi-TMDPDFs~\cite{Ebert:2018gzl,Shanahan:2019zcq,Zhang:2020dbb}.
The above factorization scheme is independent of the LF limit.

In conclusion, we presented a complete LaMET formulation for calculation of nonperturbative TMDPDFs.
We provided with the definitions of lattice calculable quasi-TMDPDFs and soft functions.
We found that the physical combination of the quasi-TMDPDFs and soft function
is related to the TMDPDFs through perturbative matching,
and the matching kernel is controlled by a renormalization group equation with known anomalous dimensions.
This allows first-principle calculations for the transverse-momentum-dependent
Drell-Yan and similar cross sections from high energy experiments.

Note that after the first version of this manuscript posted in the arXiv~\cite{Ji:2019ewn},
Ref.~\cite{Vladimirov:2020ofp} appeared, using the soft-collinear effective theory to discuss
the factorization of a similar quasi-TMDPDF in terms of physical TMDPDF, with little specifics on the
non-perturbative soft factor.

{\it Acknowledgment.}---We thank Andreas Sch\"afer, Feng Yuan, and Yong Zhao for valuable discussions.
This work is partially supported by the U.S. Department of Energy under Contract No. DE-SC0020682.

\bibliographystyle{apsrev4-1}
\bibliography{bibliography}

\begin{thebibliography}{52}%
\makeatletter
\providecommand \@ifxundefined [1]{%
 \@ifx{#1\undefined}
}%
\providecommand \@ifnum [1]{%
 \ifnum #1\expandafter \@firstoftwo
 \else \expandafter \@secondoftwo
 \fi
}%
\providecommand \@ifx [1]{%
 \ifx #1\expandafter \@firstoftwo
 \else \expandafter \@secondoftwo
 \fi
}%
\providecommand \natexlab [1]{#1}%
\providecommand \enquote  [1]{``#1''}%
\providecommand \bibnamefont  [1]{#1}%
\providecommand \bibfnamefont [1]{#1}%
\providecommand \citenamefont [1]{#1}%
\providecommand \href@noop [0]{\@secondoftwo}%
\providecommand \href [0]{\begingroup \@sanitize@url \@href}%
\providecommand \@href[1]{\@@startlink{#1}\@@href}%
\providecommand \@@href[1]{\endgroup#1\@@endlink}%
\providecommand \@sanitize@url [0]{\catcode `\\12\catcode `\$12\catcode
  `\&12\catcode `\#12\catcode `\^12\catcode `\_12\catcode `\%12\relax}%
\providecommand \@@startlink[1]{}%
\providecommand \@@endlink[0]{}%
\providecommand \url  [0]{\begingroup\@sanitize@url \@url }%
\providecommand \@url [1]{\endgroup\@href {#1}{\urlprefix }}%
\providecommand \urlprefix  [0]{URL }%
\providecommand \Eprint [0]{\href }%
\providecommand \doibase [0]{http://dx.doi.org/}%
\providecommand \selectlanguage [0]{\@gobble}%
\providecommand \bibinfo  [0]{\@secondoftwo}%
\providecommand \bibfield  [0]{\@secondoftwo}%
\providecommand \translation [1]{[#1]}%
\providecommand \BibitemOpen [0]{}%
\providecommand \bibitemStop [0]{}%
\providecommand \bibitemNoStop [0]{.\EOS\space}%
\providecommand \EOS [0]{\spacefactor3000\relax}%
\providecommand \BibitemShut  [1]{\csname bibitem#1\endcsname}%
\let\auto@bib@innerbib\@empty
\bibitem [{\citenamefont {Collins}\ and\ \citenamefont
  {Soper}(1981)}]{Collins:1981uk}%
  \BibitemOpen
  \bibfield  {author} {\bibinfo {author} {\bibfnamefont {J.~C.}\ \bibnamefont
  {Collins}}\ and\ \bibinfo {author} {\bibfnamefont {D.~E.}\ \bibnamefont
  {Soper}},\ }\href {\doibase 10.1016/0550-3213(81)90339-4} {\bibfield
  {journal} {\bibinfo  {journal} {Nucl. Phys.}\ }\textbf {\bibinfo {volume}
  {B193}},\ \bibinfo {pages} {381} (\bibinfo {year} {1981})},\ \bibinfo {note}
  {[Erratum: Nucl. Phys.B213,545(1983)]}\BibitemShut {NoStop}%
\bibitem [{\citenamefont {Collins}\ \emph {et~al.}(1985)\citenamefont
  {Collins}, \citenamefont {Soper},\ and\ \citenamefont
  {Sterman}}]{Collins:1984kg}%
  \BibitemOpen
  \bibfield  {author} {\bibinfo {author} {\bibfnamefont {J.~C.}\ \bibnamefont
  {Collins}}, \bibinfo {author} {\bibfnamefont {D.~E.}\ \bibnamefont {Soper}},
  \ and\ \bibinfo {author} {\bibfnamefont {G.~F.}\ \bibnamefont {Sterman}},\
  }\href {\doibase 10.1016/0550-3213(85)90479-1} {\bibfield  {journal}
  {\bibinfo  {journal} {Nucl. Phys.}\ }\textbf {\bibinfo {volume} {B250}},\
  \bibinfo {pages} {199} (\bibinfo {year} {1985})}\BibitemShut {NoStop}%
\bibitem [{\citenamefont {Bacchetta}\ \emph {et~al.}(2020)\citenamefont
  {Bacchetta}, \citenamefont {Bertone}, \citenamefont {Bissolotti},
  \citenamefont {Bozzi}, \citenamefont {Delcarro}, \citenamefont {Piacenza},\
  and\ \citenamefont {Radici}}]{Bacchetta:2019sam}%
  \BibitemOpen
  \bibfield  {author} {\bibinfo {author} {\bibfnamefont {A.}~\bibnamefont
  {Bacchetta}}, \bibinfo {author} {\bibfnamefont {V.}~\bibnamefont {Bertone}},
  \bibinfo {author} {\bibfnamefont {C.}~\bibnamefont {Bissolotti}}, \bibinfo
  {author} {\bibfnamefont {G.}~\bibnamefont {Bozzi}}, \bibinfo {author}
  {\bibfnamefont {F.}~\bibnamefont {Delcarro}}, \bibinfo {author}
  {\bibfnamefont {F.}~\bibnamefont {Piacenza}}, \ and\ \bibinfo {author}
  {\bibfnamefont {M.}~\bibnamefont {Radici}},\ }\href {\doibase
  10.1007/JHEP07(2020)117} {\bibfield  {journal} {\bibinfo  {journal} {JHEP}\
  }\textbf {\bibinfo {volume} {07}},\ \bibinfo {pages} {117} (\bibinfo {year}
  {2020})},\ \Eprint {http://arxiv.org/abs/1912.07550} {arXiv:1912.07550
  [hep-ph]} \BibitemShut {NoStop}%
\bibitem [{\citenamefont {Scimemi}\ and\ \citenamefont
  {Vladimirov}(2020)}]{Scimemi:2019cmh}%
  \BibitemOpen
  \bibfield  {author} {\bibinfo {author} {\bibfnamefont {I.}~\bibnamefont
  {Scimemi}}\ and\ \bibinfo {author} {\bibfnamefont {A.}~\bibnamefont
  {Vladimirov}},\ }\href {\doibase 10.1007/JHEP06(2020)137} {\bibfield
  {journal} {\bibinfo  {journal} {JHEP}\ }\textbf {\bibinfo {volume} {06}},\
  \bibinfo {pages} {137} (\bibinfo {year} {2020})},\ \Eprint
  {http://arxiv.org/abs/1912.06532} {arXiv:1912.06532 [hep-ph]} \BibitemShut
  {NoStop}%
\bibitem [{\citenamefont {Collins}\ \emph {et~al.}(1988)\citenamefont
  {Collins}, \citenamefont {Soper},\ and\ \citenamefont
  {Sterman}}]{Collins:1988ig}%
  \BibitemOpen
  \bibfield  {author} {\bibinfo {author} {\bibfnamefont {J.~C.}\ \bibnamefont
  {Collins}}, \bibinfo {author} {\bibfnamefont {D.~E.}\ \bibnamefont {Soper}},
  \ and\ \bibinfo {author} {\bibfnamefont {G.~F.}\ \bibnamefont {Sterman}},\
  }\href {\doibase 10.1016/0550-3213(88)90130-7} {\bibfield  {journal}
  {\bibinfo  {journal} {Nucl. Phys.}\ }\textbf {\bibinfo {volume} {B308}},\
  \bibinfo {pages} {833} (\bibinfo {year} {1988})}\BibitemShut {NoStop}%
\bibitem [{\citenamefont {Ji}\ \emph {et~al.}(2005)\citenamefont {Ji},
  \citenamefont {Ma},\ and\ \citenamefont {Yuan}}]{Ji:2004wu}%
  \BibitemOpen
  \bibfield  {author} {\bibinfo {author} {\bibfnamefont {X.-D.}\ \bibnamefont
  {Ji}}, \bibinfo {author} {\bibfnamefont {J.-P.}\ \bibnamefont {Ma}}, \ and\
  \bibinfo {author} {\bibfnamefont {F.}~\bibnamefont {Yuan}},\ }\href {\doibase
  10.1103/PhysRevD.71.034005} {\bibfield  {journal} {\bibinfo  {journal} {Phys.
  Rev.}\ }\textbf {\bibinfo {volume} {D71}},\ \bibinfo {pages} {034005}
  (\bibinfo {year} {2005})},\ \Eprint {http://arxiv.org/abs/hep-ph/0404183}
  {arXiv:hep-ph/0404183 [hep-ph]} \BibitemShut {NoStop}%
\bibitem [{\citenamefont {Ji}\ \emph {et~al.}(2004)\citenamefont {Ji},
  \citenamefont {Ma},\ and\ \citenamefont {Yuan}}]{Ji:2004xq}%
  \BibitemOpen
  \bibfield  {author} {\bibinfo {author} {\bibfnamefont {X.-D.}\ \bibnamefont
  {Ji}}, \bibinfo {author} {\bibfnamefont {J.-P.}\ \bibnamefont {Ma}}, \ and\
  \bibinfo {author} {\bibfnamefont {F.}~\bibnamefont {Yuan}},\ }\href {\doibase
  10.1016/j.physletb.2004.07.026} {\bibfield  {journal} {\bibinfo  {journal}
  {Phys. Lett.}\ }\textbf {\bibinfo {volume} {B597}},\ \bibinfo {pages} {299}
  (\bibinfo {year} {2004})},\ \Eprint {http://arxiv.org/abs/hep-ph/0405085}
  {arXiv:hep-ph/0405085 [hep-ph]} \BibitemShut {NoStop}%
\bibitem [{\citenamefont {Collins}(2011{\natexlab{a}})}]{Collins:2011ca}%
  \BibitemOpen
  \bibfield  {author} {\bibinfo {author} {\bibfnamefont {J.}~\bibnamefont
  {Collins}},\ }\bibfield  {booktitle} {\emph {\bibinfo {booktitle}
  {{Proceedings, QCD Evolution Workshop on From Collinear to Non-Collinear
  Case: Newport News, Virginia, April 8-9, 2011}}},\ }\href {\doibase
  10.1142/S2010194511001590} {\bibfield  {journal} {\bibinfo  {journal} {Int.
  J. Mod. Phys. Conf. Ser.}\ }\textbf {\bibinfo {volume} {4}},\ \bibinfo
  {pages} {85} (\bibinfo {year} {2011}{\natexlab{a}})},\ \Eprint
  {http://arxiv.org/abs/1107.4123} {arXiv:1107.4123 [hep-ph]} \BibitemShut
  {NoStop}%
\bibitem [{\citenamefont {Collins}\ and\ \citenamefont
  {Rogers}(2013)}]{Collins:2012uy}%
  \BibitemOpen
  \bibfield  {author} {\bibinfo {author} {\bibfnamefont {J.~C.}\ \bibnamefont
  {Collins}}\ and\ \bibinfo {author} {\bibfnamefont {T.~C.}\ \bibnamefont
  {Rogers}},\ }\href {\doibase 10.1103/PhysRevD.87.034018} {\bibfield
  {journal} {\bibinfo  {journal} {Phys. Rev.}\ }\textbf {\bibinfo {volume}
  {D87}},\ \bibinfo {pages} {034018} (\bibinfo {year} {2013})},\ \Eprint
  {http://arxiv.org/abs/1210.2100} {arXiv:1210.2100 [hep-ph]} \BibitemShut
  {NoStop}%
\bibitem [{\citenamefont {Becher}\ and\ \citenamefont
  {Neubert}(2011)}]{Becher:2010tm}%
  \BibitemOpen
  \bibfield  {author} {\bibinfo {author} {\bibfnamefont {T.}~\bibnamefont
  {Becher}}\ and\ \bibinfo {author} {\bibfnamefont {M.}~\bibnamefont
  {Neubert}},\ }\href {\doibase 10.1140/epjc/s10052-011-1665-7} {\bibfield
  {journal} {\bibinfo  {journal} {Eur. Phys. J.}\ }\textbf {\bibinfo {volume}
  {C71}},\ \bibinfo {pages} {1665} (\bibinfo {year} {2011})},\ \Eprint
  {http://arxiv.org/abs/1007.4005} {arXiv:1007.4005 [hep-ph]} \BibitemShut
  {NoStop}%
\bibitem [{\citenamefont {Echevarria}\ \emph {et~al.}(2012)\citenamefont
  {Echevarria}, \citenamefont {Idilbi},\ and\ \citenamefont
  {Scimemi}}]{GarciaEchevarria:2011rb}%
  \BibitemOpen
  \bibfield  {author} {\bibinfo {author} {\bibfnamefont {M.~G.}\ \bibnamefont
  {Echevarria}}, \bibinfo {author} {\bibfnamefont {A.}~\bibnamefont {Idilbi}},
  \ and\ \bibinfo {author} {\bibfnamefont {I.}~\bibnamefont {Scimemi}},\ }\href
  {\doibase 10.1007/JHEP07(2012)002} {\bibfield  {journal} {\bibinfo  {journal}
  {JHEP}\ }\textbf {\bibinfo {volume} {07}},\ \bibinfo {pages} {002} (\bibinfo
  {year} {2012})},\ \Eprint {http://arxiv.org/abs/1111.4996} {arXiv:1111.4996
  [hep-ph]} \BibitemShut {NoStop}%
\bibitem [{\citenamefont {Echevarría}\ \emph {et~al.}(2013)\citenamefont
  {Echevarría}, \citenamefont {Idilbi},\ and\ \citenamefont
  {Scimemi}}]{Echevarria:2012js}%
  \BibitemOpen
  \bibfield  {author} {\bibinfo {author} {\bibfnamefont {M.~G.}\ \bibnamefont
  {Echevarría}}, \bibinfo {author} {\bibfnamefont {A.}~\bibnamefont {Idilbi}},
  \ and\ \bibinfo {author} {\bibfnamefont {I.}~\bibnamefont {Scimemi}},\ }\href
  {\doibase 10.1016/j.physletb.2013.09.003} {\bibfield  {journal} {\bibinfo
  {journal} {Phys. Lett.}\ }\textbf {\bibinfo {volume} {B726}},\ \bibinfo
  {pages} {795} (\bibinfo {year} {2013})},\ \Eprint
  {http://arxiv.org/abs/1211.1947} {arXiv:1211.1947 [hep-ph]} \BibitemShut
  {NoStop}%
\bibitem [{\citenamefont {Chiu}\ \emph {et~al.}(2012)\citenamefont {Chiu},
  \citenamefont {Jain}, \citenamefont {Neill},\ and\ \citenamefont
  {Rothstein}}]{Chiu:2012ir}%
  \BibitemOpen
  \bibfield  {author} {\bibinfo {author} {\bibfnamefont {J.-Y.}\ \bibnamefont
  {Chiu}}, \bibinfo {author} {\bibfnamefont {A.}~\bibnamefont {Jain}}, \bibinfo
  {author} {\bibfnamefont {D.}~\bibnamefont {Neill}}, \ and\ \bibinfo {author}
  {\bibfnamefont {I.~Z.}\ \bibnamefont {Rothstein}},\ }\href {\doibase
  10.1007/JHEP05(2012)084} {\bibfield  {journal} {\bibinfo  {journal} {JHEP}\
  }\textbf {\bibinfo {volume} {05}},\ \bibinfo {pages} {084} (\bibinfo {year}
  {2012})},\ \Eprint {http://arxiv.org/abs/1202.0814} {arXiv:1202.0814
  [hep-ph]} \BibitemShut {NoStop}%
\bibitem [{\citenamefont {Hagler}\ \emph {et~al.}(2009)\citenamefont {Hagler},
  \citenamefont {Musch}, \citenamefont {Negele},\ and\ \citenamefont
  {Schafer}}]{Hagler:2009mb}%
  \BibitemOpen
  \bibfield  {author} {\bibinfo {author} {\bibfnamefont {P.}~\bibnamefont
  {Hagler}}, \bibinfo {author} {\bibfnamefont {B.~U.}\ \bibnamefont {Musch}},
  \bibinfo {author} {\bibfnamefont {J.~W.}\ \bibnamefont {Negele}}, \ and\
  \bibinfo {author} {\bibfnamefont {A.}~\bibnamefont {Schafer}},\ }\href
  {\doibase 10.1209/0295-5075/88/61001} {\bibfield  {journal} {\bibinfo
  {journal} {EPL}\ }\textbf {\bibinfo {volume} {88}},\ \bibinfo {pages} {61001}
  (\bibinfo {year} {2009})},\ \Eprint {http://arxiv.org/abs/0908.1283}
  {arXiv:0908.1283 [hep-lat]} \BibitemShut {NoStop}%
\bibitem [{\citenamefont {Musch}\ \emph {et~al.}(2011)\citenamefont {Musch},
  \citenamefont {Hagler}, \citenamefont {Negele},\ and\ \citenamefont
  {Schafer}}]{Musch:2010ka}%
  \BibitemOpen
  \bibfield  {author} {\bibinfo {author} {\bibfnamefont {B.~U.}\ \bibnamefont
  {Musch}}, \bibinfo {author} {\bibfnamefont {P.}~\bibnamefont {Hagler}},
  \bibinfo {author} {\bibfnamefont {J.~W.}\ \bibnamefont {Negele}}, \ and\
  \bibinfo {author} {\bibfnamefont {A.}~\bibnamefont {Schafer}},\ }\href
  {\doibase 10.1103/PhysRevD.83.094507} {\bibfield  {journal} {\bibinfo
  {journal} {Phys. Rev. D}\ }\textbf {\bibinfo {volume} {83}},\ \bibinfo
  {pages} {094507} (\bibinfo {year} {2011})},\ \Eprint
  {http://arxiv.org/abs/1011.1213} {arXiv:1011.1213 [hep-lat]} \BibitemShut
  {NoStop}%
\bibitem [{\citenamefont {Musch}\ \emph {et~al.}(2012)\citenamefont {Musch},
  \citenamefont {Hagler}, \citenamefont {Engelhardt}, \citenamefont {Negele},\
  and\ \citenamefont {Schafer}}]{Musch:2011er}%
  \BibitemOpen
  \bibfield  {author} {\bibinfo {author} {\bibfnamefont {B.~U.}\ \bibnamefont
  {Musch}}, \bibinfo {author} {\bibfnamefont {P.}~\bibnamefont {Hagler}},
  \bibinfo {author} {\bibfnamefont {M.}~\bibnamefont {Engelhardt}}, \bibinfo
  {author} {\bibfnamefont {J.~W.}\ \bibnamefont {Negele}}, \ and\ \bibinfo
  {author} {\bibfnamefont {A.}~\bibnamefont {Schafer}},\ }\href {\doibase
  10.1103/PhysRevD.85.094510} {\bibfield  {journal} {\bibinfo  {journal} {Phys.
  Rev.}\ }\textbf {\bibinfo {volume} {D85}},\ \bibinfo {pages} {094510}
  (\bibinfo {year} {2012})},\ \Eprint {http://arxiv.org/abs/1111.4249}
  {arXiv:1111.4249 [hep-lat]} \BibitemShut {NoStop}%
\bibitem [{\citenamefont {Engelhardt}\ \emph {et~al.}(2016)\citenamefont
  {Engelhardt}, \citenamefont {Hägler}, \citenamefont {Musch}, \citenamefont
  {Negele},\ and\ \citenamefont {Schäfer}}]{Engelhardt:2015xja}%
  \BibitemOpen
  \bibfield  {author} {\bibinfo {author} {\bibfnamefont {M.}~\bibnamefont
  {Engelhardt}}, \bibinfo {author} {\bibfnamefont {P.}~\bibnamefont {Hägler}},
  \bibinfo {author} {\bibfnamefont {B.}~\bibnamefont {Musch}}, \bibinfo
  {author} {\bibfnamefont {J.}~\bibnamefont {Negele}}, \ and\ \bibinfo {author}
  {\bibfnamefont {A.}~\bibnamefont {Schäfer}},\ }\href {\doibase
  10.1103/PhysRevD.93.054501} {\bibfield  {journal} {\bibinfo  {journal} {Phys.
  Rev.}\ }\textbf {\bibinfo {volume} {D93}},\ \bibinfo {pages} {054501}
  (\bibinfo {year} {2016})},\ \Eprint {http://arxiv.org/abs/1506.07826}
  {arXiv:1506.07826 [hep-lat]} \BibitemShut {NoStop}%
\bibitem [{\citenamefont {Yoon}\ \emph {et~al.}(2017)\citenamefont {Yoon},
  \citenamefont {Engelhardt}, \citenamefont {Gupta}, \citenamefont
  {Bhattacharya}, \citenamefont {Green}, \citenamefont {Musch}, \citenamefont
  {Negele}, \citenamefont {Pochinsky}, \citenamefont {Schäfer},\ and\
  \citenamefont {Syritsyn}}]{Yoon:2017qzo}%
  \BibitemOpen
  \bibfield  {author} {\bibinfo {author} {\bibfnamefont {B.}~\bibnamefont
  {Yoon}}, \bibinfo {author} {\bibfnamefont {M.}~\bibnamefont {Engelhardt}},
  \bibinfo {author} {\bibfnamefont {R.}~\bibnamefont {Gupta}}, \bibinfo
  {author} {\bibfnamefont {T.}~\bibnamefont {Bhattacharya}}, \bibinfo {author}
  {\bibfnamefont {J.~R.}\ \bibnamefont {Green}}, \bibinfo {author}
  {\bibfnamefont {B.~U.}\ \bibnamefont {Musch}}, \bibinfo {author}
  {\bibfnamefont {J.~W.}\ \bibnamefont {Negele}}, \bibinfo {author}
  {\bibfnamefont {A.~V.}\ \bibnamefont {Pochinsky}}, \bibinfo {author}
  {\bibfnamefont {A.}~\bibnamefont {Schäfer}}, \ and\ \bibinfo {author}
  {\bibfnamefont {S.~N.}\ \bibnamefont {Syritsyn}},\ }\href {\doibase
  10.1103/PhysRevD.96.094508} {\bibfield  {journal} {\bibinfo  {journal} {Phys.
  Rev.}\ }\textbf {\bibinfo {volume} {D96}},\ \bibinfo {pages} {094508}
  (\bibinfo {year} {2017})},\ \Eprint {http://arxiv.org/abs/1706.03406}
  {arXiv:1706.03406 [hep-lat]} \BibitemShut {NoStop}%
\bibitem [{\citenamefont {Ji}(2013)}]{Ji:2013dva}%
  \BibitemOpen
  \bibfield  {author} {\bibinfo {author} {\bibfnamefont {X.}~\bibnamefont
  {Ji}},\ }\href {\doibase 10.1103/PhysRevLett.110.262002} {\bibfield
  {journal} {\bibinfo  {journal} {Phys. Rev. Lett.}\ }\textbf {\bibinfo
  {volume} {110}},\ \bibinfo {pages} {262002} (\bibinfo {year} {2013})},\
  \Eprint {http://arxiv.org/abs/1305.1539} {arXiv:1305.1539 [hep-ph]}
  \BibitemShut {NoStop}%
\bibitem [{\citenamefont {Ji}(2014)}]{Ji:2014gla}%
  \BibitemOpen
  \bibfield  {author} {\bibinfo {author} {\bibfnamefont {X.}~\bibnamefont
  {Ji}},\ }\href {\doibase 10.1007/s11433-014-5492-3} {\bibfield  {journal}
  {\bibinfo  {journal} {Sci. China Phys. Mech. Astron.}\ }\textbf {\bibinfo
  {volume} {57}},\ \bibinfo {pages} {1407} (\bibinfo {year} {2014})},\ \Eprint
  {http://arxiv.org/abs/1404.6680} {arXiv:1404.6680 [hep-ph]} \BibitemShut
  {NoStop}%
\bibitem [{\citenamefont {Ji}\ \emph {et~al.}(2015)\citenamefont {Ji},
  \citenamefont {Sun}, \citenamefont {Xiong},\ and\ \citenamefont
  {Yuan}}]{Ji:2014hxa}%
  \BibitemOpen
  \bibfield  {author} {\bibinfo {author} {\bibfnamefont {X.}~\bibnamefont
  {Ji}}, \bibinfo {author} {\bibfnamefont {P.}~\bibnamefont {Sun}}, \bibinfo
  {author} {\bibfnamefont {X.}~\bibnamefont {Xiong}}, \ and\ \bibinfo {author}
  {\bibfnamefont {F.}~\bibnamefont {Yuan}},\ }\href {\doibase
  10.1103/PhysRevD.91.074009} {\bibfield  {journal} {\bibinfo  {journal} {Phys.
  Rev.}\ }\textbf {\bibinfo {volume} {D91}},\ \bibinfo {pages} {074009}
  (\bibinfo {year} {2015})},\ \Eprint {http://arxiv.org/abs/1405.7640}
  {arXiv:1405.7640 [hep-ph]} \BibitemShut {NoStop}%
\bibitem [{\citenamefont {Ji}\ \emph {et~al.}(2019{\natexlab{a}})\citenamefont
  {Ji}, \citenamefont {Jin}, \citenamefont {Yuan}, \citenamefont {Zhang},\ and\
  \citenamefont {Zhao}}]{Ji:2018hvs}%
  \BibitemOpen
  \bibfield  {author} {\bibinfo {author} {\bibfnamefont {X.}~\bibnamefont
  {Ji}}, \bibinfo {author} {\bibfnamefont {L.-C.}\ \bibnamefont {Jin}},
  \bibinfo {author} {\bibfnamefont {F.}~\bibnamefont {Yuan}}, \bibinfo {author}
  {\bibfnamefont {J.-H.}\ \bibnamefont {Zhang}}, \ and\ \bibinfo {author}
  {\bibfnamefont {Y.}~\bibnamefont {Zhao}},\ }\href {\doibase
  10.1103/PhysRevD.99.114006} {\bibfield  {journal} {\bibinfo  {journal} {Phys.
  Rev.}\ }\textbf {\bibinfo {volume} {D99}},\ \bibinfo {pages} {114006}
  (\bibinfo {year} {2019}{\natexlab{a}})},\ \Eprint
  {http://arxiv.org/abs/1801.05930} {arXiv:1801.05930 [hep-ph]} \BibitemShut
  {NoStop}%
\bibitem [{\citenamefont {Ebert}\ \emph
  {et~al.}(2019{\natexlab{a}})\citenamefont {Ebert}, \citenamefont {Stewart},\
  and\ \citenamefont {Zhao}}]{Ebert:2018gzl}%
  \BibitemOpen
  \bibfield  {author} {\bibinfo {author} {\bibfnamefont {M.~A.}\ \bibnamefont
  {Ebert}}, \bibinfo {author} {\bibfnamefont {I.~W.}\ \bibnamefont {Stewart}},
  \ and\ \bibinfo {author} {\bibfnamefont {Y.}~\bibnamefont {Zhao}},\ }\href
  {\doibase 10.1103/PhysRevD.99.034505} {\bibfield  {journal} {\bibinfo
  {journal} {Phys. Rev.}\ }\textbf {\bibinfo {volume} {D99}},\ \bibinfo {pages}
  {034505} (\bibinfo {year} {2019}{\natexlab{a}})},\ \Eprint
  {http://arxiv.org/abs/1811.00026} {arXiv:1811.00026 [hep-ph]} \BibitemShut
  {NoStop}%
\bibitem [{\citenamefont {Ebert}\ \emph
  {et~al.}(2019{\natexlab{b}})\citenamefont {Ebert}, \citenamefont {Stewart},\
  and\ \citenamefont {Zhao}}]{Ebert:2019okf}%
  \BibitemOpen
  \bibfield  {author} {\bibinfo {author} {\bibfnamefont {M.~A.}\ \bibnamefont
  {Ebert}}, \bibinfo {author} {\bibfnamefont {I.~W.}\ \bibnamefont {Stewart}},
  \ and\ \bibinfo {author} {\bibfnamefont {Y.}~\bibnamefont {Zhao}},\ }\href
  {\doibase 10.1007/JHEP09(2019)037} {\bibfield  {journal} {\bibinfo  {journal}
  {JHEP}\ }\textbf {\bibinfo {volume} {09}},\ \bibinfo {pages} {037} (\bibinfo
  {year} {2019}{\natexlab{b}})},\ \Eprint {http://arxiv.org/abs/1901.03685}
  {arXiv:1901.03685 [hep-ph]} \BibitemShut {NoStop}%
\bibitem [{\citenamefont {Ji}\ \emph {et~al.}(2019{\natexlab{b}})\citenamefont
  {Ji}, \citenamefont {Liu},\ and\ \citenamefont {Liu}}]{Ji:2019sxk}%
  \BibitemOpen
  \bibfield  {author} {\bibinfo {author} {\bibfnamefont {X.}~\bibnamefont
  {Ji}}, \bibinfo {author} {\bibfnamefont {Y.}~\bibnamefont {Liu}}, \ and\
  \bibinfo {author} {\bibfnamefont {Y.-S.}\ \bibnamefont {Liu}},\ }\href@noop
  {} {\  (\bibinfo {year} {2019}{\natexlab{b}})},\ \Eprint
  {http://arxiv.org/abs/1910.11415} {arXiv:1910.11415 [hep-ph]} \BibitemShut
  {NoStop}%
\bibitem [{\citenamefont {Collins}(2011{\natexlab{b}})}]{Collins:2011zzd}%
  \BibitemOpen
  \bibfield  {author} {\bibinfo {author} {\bibfnamefont {J.}~\bibnamefont
  {Collins}},\ }\href@noop {} {\bibfield  {journal} {\bibinfo  {journal} {Camb.
  Monogr. Part. Phys. Nucl. Phys. Cosmol.}\ }\textbf {\bibinfo {volume} {32}},\
  \bibinfo {pages} {1} (\bibinfo {year} {2011}{\natexlab{b}})}\BibitemShut
  {NoStop}%
\bibitem [{\citenamefont {Belitsky}\ \emph {et~al.}(2003)\citenamefont
  {Belitsky}, \citenamefont {Ji},\ and\ \citenamefont
  {Yuan}}]{Belitsky:2002sm}%
  \BibitemOpen
  \bibfield  {author} {\bibinfo {author} {\bibfnamefont {A.~V.}\ \bibnamefont
  {Belitsky}}, \bibinfo {author} {\bibfnamefont {X.}~\bibnamefont {Ji}}, \ and\
  \bibinfo {author} {\bibfnamefont {F.}~\bibnamefont {Yuan}},\ }\href {\doibase
  10.1016/S0550-3213(03)00121-4} {\bibfield  {journal} {\bibinfo  {journal}
  {Nucl. Phys. B}\ }\textbf {\bibinfo {volume} {656}},\ \bibinfo {pages} {165}
  (\bibinfo {year} {2003})},\ \Eprint {http://arxiv.org/abs/hep-ph/0208038}
  {arXiv:hep-ph/0208038} \BibitemShut {NoStop}%
\bibitem [{\citenamefont {Collins}\ and\ \citenamefont
  {Metz}(2004)}]{Collins:2004nx}%
  \BibitemOpen
  \bibfield  {author} {\bibinfo {author} {\bibfnamefont {J.~C.}\ \bibnamefont
  {Collins}}\ and\ \bibinfo {author} {\bibfnamefont {A.}~\bibnamefont {Metz}},\
  }\href {\doibase 10.1103/PhysRevLett.93.252001} {\bibfield  {journal}
  {\bibinfo  {journal} {Phys. Rev. Lett.}\ }\textbf {\bibinfo {volume} {93}},\
  \bibinfo {pages} {252001} (\bibinfo {year} {2004})},\ \Eprint
  {http://arxiv.org/abs/hep-ph/0408249} {arXiv:hep-ph/0408249 [hep-ph]}
  \BibitemShut {NoStop}%
\bibitem [{\citenamefont {Korchemsky}\ and\ \citenamefont
  {Radyushkin}(1987)}]{Korchemsky:1987wg}%
  \BibitemOpen
  \bibfield  {author} {\bibinfo {author} {\bibfnamefont {G.~P.}\ \bibnamefont
  {Korchemsky}}\ and\ \bibinfo {author} {\bibfnamefont {A.~V.}\ \bibnamefont
  {Radyushkin}},\ }\href {\doibase 10.1016/0550-3213(87)90277-X} {\bibfield
  {journal} {\bibinfo  {journal} {Nucl. Phys.}\ }\textbf {\bibinfo {volume}
  {B283}},\ \bibinfo {pages} {342} (\bibinfo {year} {1987})}\BibitemShut
  {NoStop}%
\bibitem [{\citenamefont {Grozin}\ \emph {et~al.}(2016)\citenamefont {Grozin},
  \citenamefont {Henn}, \citenamefont {Korchemsky},\ and\ \citenamefont
  {Marquard}}]{Grozin:2015kna}%
  \BibitemOpen
  \bibfield  {author} {\bibinfo {author} {\bibfnamefont {A.}~\bibnamefont
  {Grozin}}, \bibinfo {author} {\bibfnamefont {J.~M.}\ \bibnamefont {Henn}},
  \bibinfo {author} {\bibfnamefont {G.~P.}\ \bibnamefont {Korchemsky}}, \ and\
  \bibinfo {author} {\bibfnamefont {P.}~\bibnamefont {Marquard}},\ }\href
  {\doibase 10.1007/JHEP01(2016)140} {\bibfield  {journal} {\bibinfo  {journal}
  {JHEP}\ }\textbf {\bibinfo {volume} {01}},\ \bibinfo {pages} {140} (\bibinfo
  {year} {2016})},\ \Eprint {http://arxiv.org/abs/1510.07803} {arXiv:1510.07803
  [hep-ph]} \BibitemShut {NoStop}%
\bibitem [{\citenamefont {Ji}(2020)}]{Ji:2020byp}%
  \BibitemOpen
  \bibfield  {author} {\bibinfo {author} {\bibfnamefont {X.}~\bibnamefont
  {Ji}},\ }\href@noop {} {\  (\bibinfo {year} {2020})},\ \Eprint
  {http://arxiv.org/abs/2007.06613} {arXiv:2007.06613 [hep-ph]} \BibitemShut
  {NoStop}%
\bibitem [{\citenamefont {Feynman}(1973)}]{Feynman:1973xc}%
  \BibitemOpen
  \bibfield  {author} {\bibinfo {author} {\bibfnamefont {R.}~\bibnamefont
  {Feynman}},\ }\href@noop {} {\  (\bibinfo {year} {1973})}\BibitemShut
  {NoStop}%
\bibitem [{\citenamefont {Hatta}\ \emph {et~al.}(2014)\citenamefont {Hatta},
  \citenamefont {Ji},\ and\ \citenamefont {Zhao}}]{Hatta:2013gta}%
  \BibitemOpen
  \bibfield  {author} {\bibinfo {author} {\bibfnamefont {Y.}~\bibnamefont
  {Hatta}}, \bibinfo {author} {\bibfnamefont {X.}~\bibnamefont {Ji}}, \ and\
  \bibinfo {author} {\bibfnamefont {Y.}~\bibnamefont {Zhao}},\ }\href {\doibase
  10.1103/PhysRevD.89.085030} {\bibfield  {journal} {\bibinfo  {journal} {Phys.
  Rev. D}\ }\textbf {\bibinfo {volume} {89}},\ \bibinfo {pages} {085030}
  (\bibinfo {year} {2014})},\ \Eprint {http://arxiv.org/abs/1310.4263}
  {arXiv:1310.4263 [hep-ph]} \BibitemShut {NoStop}%
\bibitem [{\citenamefont {Manohar}\ and\ \citenamefont
  {Wise}(2000)}]{Manohar:2000dt}%
  \BibitemOpen
  \bibfield  {author} {\bibinfo {author} {\bibfnamefont {A.~V.}\ \bibnamefont
  {Manohar}}\ and\ \bibinfo {author} {\bibfnamefont {M.~B.}\ \bibnamefont
  {Wise}},\ }\href@noop {} {\bibfield  {journal} {\bibinfo  {journal} {Camb.
  Monogr. Part. Phys. Nucl. Phys. Cosmol.}\ }\textbf {\bibinfo {volume} {10}},\
  \bibinfo {pages} {1} (\bibinfo {year} {2000})}\BibitemShut {NoStop}%
\bibitem [{\citenamefont {Collins}(2008)}]{Collins:2008ht}%
  \BibitemOpen
  \bibfield  {author} {\bibinfo {author} {\bibfnamefont {J.}~\bibnamefont
  {Collins}},\ }\bibfield  {booktitle} {\emph {\bibinfo {booktitle}
  {{Proceedings, International Workshop on Relativistic nuclear and particle
  physics (Light Cone 2008): Mulhouse, France, July 7-11, 2008}}},\ }\href
  {\doibase 10.22323/1.061.0028} {\bibfield  {journal} {\bibinfo  {journal}
  {PoS}\ }\textbf {\bibinfo {volume} {LC2008}},\ \bibinfo {pages} {028}
  (\bibinfo {year} {2008})},\ \Eprint {http://arxiv.org/abs/0808.2665}
  {arXiv:0808.2665 [hep-ph]} \BibitemShut {NoStop}%
\bibitem [{\citenamefont {Ji}\ \emph {et~al.}(2020)\citenamefont {Ji},
  \citenamefont {Liu}, \citenamefont {Liu}, \citenamefont {Zhang},\ and\
  \citenamefont {Zhao}}]{Ji:2020ect}%
  \BibitemOpen
  \bibfield  {author} {\bibinfo {author} {\bibfnamefont {X.}~\bibnamefont
  {Ji}}, \bibinfo {author} {\bibfnamefont {Y.-S.}\ \bibnamefont {Liu}},
  \bibinfo {author} {\bibfnamefont {Y.}~\bibnamefont {Liu}}, \bibinfo {author}
  {\bibfnamefont {J.-H.}\ \bibnamefont {Zhang}}, \ and\ \bibinfo {author}
  {\bibfnamefont {Y.}~\bibnamefont {Zhao}},\ }\href@noop {} {\  (\bibinfo
  {year} {2020})},\ \Eprint {http://arxiv.org/abs/2004.03543} {arXiv:2004.03543
  [hep-ph]} \BibitemShut {NoStop}%
\bibitem [{\citenamefont {Shanahan}\ \emph {et~al.}(2019)\citenamefont
  {Shanahan}, \citenamefont {Wagman},\ and\ \citenamefont
  {Zhao}}]{Shanahan:2019zcq}%
  \BibitemOpen
  \bibfield  {author} {\bibinfo {author} {\bibfnamefont {P.}~\bibnamefont
  {Shanahan}}, \bibinfo {author} {\bibfnamefont {M.}~\bibnamefont {Wagman}}, \
  and\ \bibinfo {author} {\bibfnamefont {Y.}~\bibnamefont {Zhao}},\ }\href@noop
  {} {\  (\bibinfo {year} {2019})},\ \Eprint {http://arxiv.org/abs/1911.00800}
  {arXiv:1911.00800 [hep-lat]} \BibitemShut {NoStop}%
\bibitem [{\citenamefont {Zhang}\ \emph {et~al.}(2020)\citenamefont {Zhang}
  \emph {et~al.}}]{Zhang:2020dbb}%
  \BibitemOpen
  \bibfield  {author} {\bibinfo {author} {\bibfnamefont {Q.-A.}\ \bibnamefont
  {Zhang}} \emph {et~al.} (\bibinfo {collaboration} {Lattice Parton}),\
  }\href@noop {} {\  (\bibinfo {year} {2020})},\ \Eprint
  {http://arxiv.org/abs/2005.14572} {arXiv:2005.14572 [hep-lat]} \BibitemShut
  {NoStop}%
\bibitem [{\citenamefont {Falk}\ \emph {et~al.}(1990)\citenamefont {Falk},
  \citenamefont {Georgi}, \citenamefont {Grinstein},\ and\ \citenamefont
  {Wise}}]{Falk:1990yz}%
  \BibitemOpen
  \bibfield  {author} {\bibinfo {author} {\bibfnamefont {A.~F.}\ \bibnamefont
  {Falk}}, \bibinfo {author} {\bibfnamefont {H.}~\bibnamefont {Georgi}},
  \bibinfo {author} {\bibfnamefont {B.}~\bibnamefont {Grinstein}}, \ and\
  \bibinfo {author} {\bibfnamefont {M.~B.}\ \bibnamefont {Wise}},\ }\href
  {\doibase 10.1016/0550-3213(90)90591-Z} {\bibfield  {journal} {\bibinfo
  {journal} {Nucl. Phys. B}\ }\textbf {\bibinfo {volume} {343}},\ \bibinfo
  {pages} {1} (\bibinfo {year} {1990})}\BibitemShut {NoStop}%
\bibitem [{\citenamefont {Ji}\ and\ \citenamefont {Musolf}(1991)}]{Ji:1991pr}%
  \BibitemOpen
  \bibfield  {author} {\bibinfo {author} {\bibfnamefont {X.-D.}\ \bibnamefont
  {Ji}}\ and\ \bibinfo {author} {\bibfnamefont {M.~J.}\ \bibnamefont
  {Musolf}},\ }\href {\doibase 10.1016/0370-2693(91)91916-J} {\bibfield
  {journal} {\bibinfo  {journal} {Phys. Lett.}\ }\textbf {\bibinfo {volume}
  {B257}},\ \bibinfo {pages} {409} (\bibinfo {year} {1991})}\BibitemShut
  {NoStop}%
\bibitem [{\citenamefont {Braun}\ \emph {et~al.}(2020)\citenamefont {Braun},
  \citenamefont {Chetyrkin},\ and\ \citenamefont {Kniehl}}]{Braun:2020ymy}%
  \BibitemOpen
  \bibfield  {author} {\bibinfo {author} {\bibfnamefont {V.}~\bibnamefont
  {Braun}}, \bibinfo {author} {\bibfnamefont {K.}~\bibnamefont {Chetyrkin}}, \
  and\ \bibinfo {author} {\bibfnamefont {B.}~\bibnamefont {Kniehl}},\ }\href
  {\doibase 10.1007/JHEP07(2020)161} {\bibfield  {journal} {\bibinfo  {journal}
  {JHEP}\ }\textbf {\bibinfo {volume} {07}},\ \bibinfo {pages} {161} (\bibinfo
  {year} {2020})},\ \Eprint {http://arxiv.org/abs/2004.01043} {arXiv:2004.01043
  [hep-ph]} \BibitemShut {NoStop}%
\bibitem [{\citenamefont {Ji}\ \emph {et~al.}(pear)\citenamefont {Ji},
  \citenamefont {Liu},\ and\ \citenamefont {Liu}}]{future}%
  \BibitemOpen
  \bibfield  {author} {\bibinfo {author} {\bibfnamefont {X.}~\bibnamefont
  {Ji}}, \bibinfo {author} {\bibfnamefont {Y.}~\bibnamefont {Liu}}, \ and\
  \bibinfo {author} {\bibfnamefont {Y.-S.}\ \bibnamefont {Liu}},\ }\href@noop
  {} {\  (\bibinfo {year} {To appear})}\BibitemShut {NoStop}%
\bibitem [{\citenamefont {Chetyrkin}\ and\ \citenamefont
  {Grozin}(2003)}]{Chetyrkin:2003vi}%
  \BibitemOpen
  \bibfield  {author} {\bibinfo {author} {\bibfnamefont {K.~G.}\ \bibnamefont
  {Chetyrkin}}\ and\ \bibinfo {author} {\bibfnamefont {A.~G.}\ \bibnamefont
  {Grozin}},\ }\href {\doibase 10.1016/S0550-3213(03)00490-5} {\bibfield
  {journal} {\bibinfo  {journal} {Nucl. Phys.}\ }\textbf {\bibinfo {volume}
  {B666}},\ \bibinfo {pages} {289} (\bibinfo {year} {2003})},\ \Eprint
  {http://arxiv.org/abs/hep-ph/0303113} {arXiv:hep-ph/0303113 [hep-ph]}
  \BibitemShut {NoStop}%
\bibitem [{\citenamefont {Luo}\ \emph {et~al.}(2019)\citenamefont {Luo},
  \citenamefont {Wang}, \citenamefont {Xu}, \citenamefont {Yang}, \citenamefont
  {Yang},\ and\ \citenamefont {Zhu}}]{Luo:2019hmp}%
  \BibitemOpen
  \bibfield  {author} {\bibinfo {author} {\bibfnamefont {M.-X.}\ \bibnamefont
  {Luo}}, \bibinfo {author} {\bibfnamefont {X.}~\bibnamefont {Wang}}, \bibinfo
  {author} {\bibfnamefont {X.}~\bibnamefont {Xu}}, \bibinfo {author}
  {\bibfnamefont {L.~L.}\ \bibnamefont {Yang}}, \bibinfo {author}
  {\bibfnamefont {T.-Z.}\ \bibnamefont {Yang}}, \ and\ \bibinfo {author}
  {\bibfnamefont {H.~X.}\ \bibnamefont {Zhu}},\ }\href {\doibase
  10.1007/JHEP10(2019)083} {\bibfield  {journal} {\bibinfo  {journal} {JHEP}\
  }\textbf {\bibinfo {volume} {10}},\ \bibinfo {pages} {083} (\bibinfo {year}
  {2019})},\ \Eprint {http://arxiv.org/abs/1908.03831} {arXiv:1908.03831
  [hep-ph]} \BibitemShut {NoStop}%
\bibitem [{\citenamefont {Dotsenko}\ and\ \citenamefont
  {Vergeles}(1980)}]{Dotsenko:1979wb}%
  \BibitemOpen
  \bibfield  {author} {\bibinfo {author} {\bibfnamefont {V.~S.}\ \bibnamefont
  {Dotsenko}}\ and\ \bibinfo {author} {\bibfnamefont {S.~N.}\ \bibnamefont
  {Vergeles}},\ }\href {\doibase 10.1016/0550-3213(80)90103-0} {\bibfield
  {journal} {\bibinfo  {journal} {Nucl. Phys.}\ }\textbf {\bibinfo {volume}
  {B169}},\ \bibinfo {pages} {527} (\bibinfo {year} {1980})}\BibitemShut
  {NoStop}%
\bibitem [{\citenamefont {Brandt}\ \emph {et~al.}(1981)\citenamefont {Brandt},
  \citenamefont {Neri},\ and\ \citenamefont {Sato}}]{Brandt:1981kf}%
  \BibitemOpen
  \bibfield  {author} {\bibinfo {author} {\bibfnamefont {R.~A.}\ \bibnamefont
  {Brandt}}, \bibinfo {author} {\bibfnamefont {F.}~\bibnamefont {Neri}}, \ and\
  \bibinfo {author} {\bibfnamefont {M.-a.}\ \bibnamefont {Sato}},\ }\href
  {\doibase 10.1103/PhysRevD.24.879} {\bibfield  {journal} {\bibinfo  {journal}
  {Phys. Rev. D}\ }\textbf {\bibinfo {volume} {24}},\ \bibinfo {pages} {879}
  (\bibinfo {year} {1981})}\BibitemShut {NoStop}%
\bibitem [{\citenamefont {Ishikawa}\ \emph {et~al.}(2017)\citenamefont
  {Ishikawa}, \citenamefont {Ma}, \citenamefont {Qiu},\ and\ \citenamefont
  {Yoshida}}]{Ishikawa:2017faj}%
  \BibitemOpen
  \bibfield  {author} {\bibinfo {author} {\bibfnamefont {T.}~\bibnamefont
  {Ishikawa}}, \bibinfo {author} {\bibfnamefont {Y.-Q.}\ \bibnamefont {Ma}},
  \bibinfo {author} {\bibfnamefont {J.-W.}\ \bibnamefont {Qiu}}, \ and\
  \bibinfo {author} {\bibfnamefont {S.}~\bibnamefont {Yoshida}},\ }\href
  {\doibase 10.1103/PhysRevD.96.094019} {\bibfield  {journal} {\bibinfo
  {journal} {Phys. Rev.}\ }\textbf {\bibinfo {volume} {D96}},\ \bibinfo {pages}
  {094019} (\bibinfo {year} {2017})},\ \Eprint
  {http://arxiv.org/abs/1707.03107} {arXiv:1707.03107 [hep-ph]} \BibitemShut
  {NoStop}%
\bibitem [{\citenamefont {Ji}\ \emph {et~al.}(2018)\citenamefont {Ji},
  \citenamefont {Zhang},\ and\ \citenamefont {Zhao}}]{Ji:2017oey}%
  \BibitemOpen
  \bibfield  {author} {\bibinfo {author} {\bibfnamefont {X.}~\bibnamefont
  {Ji}}, \bibinfo {author} {\bibfnamefont {J.-H.}\ \bibnamefont {Zhang}}, \
  and\ \bibinfo {author} {\bibfnamefont {Y.}~\bibnamefont {Zhao}},\ }\href
  {\doibase 10.1103/PhysRevLett.120.112001} {\bibfield  {journal} {\bibinfo
  {journal} {Phys. Rev. Lett.}\ }\textbf {\bibinfo {volume} {120}},\ \bibinfo
  {pages} {112001} (\bibinfo {year} {2018})},\ \Eprint
  {http://arxiv.org/abs/1706.08962} {arXiv:1706.08962 [hep-ph]} \BibitemShut
  {NoStop}%
\bibitem [{\citenamefont {Moch}\ \emph {et~al.}(2005)\citenamefont {Moch},
  \citenamefont {Vermaseren},\ and\ \citenamefont {Vogt}}]{Moch:2005tm}%
  \BibitemOpen
  \bibfield  {author} {\bibinfo {author} {\bibfnamefont {S.}~\bibnamefont
  {Moch}}, \bibinfo {author} {\bibfnamefont {J.}~\bibnamefont {Vermaseren}}, \
  and\ \bibinfo {author} {\bibfnamefont {A.}~\bibnamefont {Vogt}},\ }\href
  {\doibase 10.1016/j.physletb.2005.08.067} {\bibfield  {journal} {\bibinfo
  {journal} {Phys. Lett. B}\ }\textbf {\bibinfo {volume} {625}},\ \bibinfo
  {pages} {245} (\bibinfo {year} {2005})},\ \Eprint
  {http://arxiv.org/abs/hep-ph/0508055} {arXiv:hep-ph/0508055} \BibitemShut
  {NoStop}%
\bibitem [{\citenamefont {Baikov}\ \emph {et~al.}(2009)\citenamefont {Baikov},
  \citenamefont {Chetyrkin}, \citenamefont {Smirnov}, \citenamefont {Smirnov},\
  and\ \citenamefont {Steinhauser}}]{Baikov:2009bg}%
  \BibitemOpen
  \bibfield  {author} {\bibinfo {author} {\bibfnamefont {P.}~\bibnamefont
  {Baikov}}, \bibinfo {author} {\bibfnamefont {K.}~\bibnamefont {Chetyrkin}},
  \bibinfo {author} {\bibfnamefont {A.}~\bibnamefont {Smirnov}}, \bibinfo
  {author} {\bibfnamefont {V.}~\bibnamefont {Smirnov}}, \ and\ \bibinfo
  {author} {\bibfnamefont {M.}~\bibnamefont {Steinhauser}},\ }\href {\doibase
  10.1103/PhysRevLett.102.212002} {\bibfield  {journal} {\bibinfo  {journal}
  {Phys. Rev. Lett.}\ }\textbf {\bibinfo {volume} {102}},\ \bibinfo {pages}
  {212002} (\bibinfo {year} {2009})},\ \Eprint {http://arxiv.org/abs/0902.3519}
  {arXiv:0902.3519 [hep-ph]} \BibitemShut {NoStop}%
\bibitem [{\citenamefont {Ji}\ \emph {et~al.}(2019{\natexlab{c}})\citenamefont
  {Ji}, \citenamefont {Liu},\ and\ \citenamefont {Liu}}]{Ji:2019ewn}%
  \BibitemOpen
  \bibfield  {author} {\bibinfo {author} {\bibfnamefont {X.}~\bibnamefont
  {Ji}}, \bibinfo {author} {\bibfnamefont {Y.}~\bibnamefont {Liu}}, \ and\
  \bibinfo {author} {\bibfnamefont {Y.-S.}\ \bibnamefont {Liu}},\ }\href@noop
  {} {\  (\bibinfo {year} {2019}{\natexlab{c}})},\ \Eprint
  {http://arxiv.org/abs/1911.03840} {arXiv:1911.03840 [hep-ph]} \BibitemShut
  {NoStop}%
\bibitem [{\citenamefont {Vladimirov}\ and\ \citenamefont
  {Schäfer}(2020)}]{Vladimirov:2020ofp}%
  \BibitemOpen
  \bibfield  {author} {\bibinfo {author} {\bibfnamefont {A.~A.}\ \bibnamefont
  {Vladimirov}}\ and\ \bibinfo {author} {\bibfnamefont {A.}~\bibnamefont
  {Schäfer}},\ }\href {\doibase 10.1103/PhysRevD.101.074517} {\bibfield
  {journal} {\bibinfo  {journal} {Phys. Rev. D}\ }\textbf {\bibinfo {volume}
  {101}},\ \bibinfo {pages} {074517} (\bibinfo {year} {2020})},\ \Eprint
  {http://arxiv.org/abs/2002.07527} {arXiv:2002.07527 [hep-ph]} \BibitemShut
  {NoStop}%
\end{thebibliography}%

\end{document}